\documentclass{aastex}

\usepackage{graphicx}
\usepackage{emulateapj5}
\usepackage{times}



\newcommand{\lsim}{\mbox{\hspace{.2em}\raisebox{.5ex}{$<$}\hspace{-.8em}\raisebox{-.5ex}{$\sim$}\hspace{.2em}}}

\def\chandra    {{\em Chandra}\/}

\def\einstein   {{\em Einstein}\/}

\def\rosat      {{\em ROSAT}\/}

\def\hst        {{\em HST}\/}

\usepackage{mathptmx}

\begin{document}


\title{Revealing the interaction between the X-ray gas of starburst galaxy UGC~6697
and the hot intracluster medium of A1367}

\author{M.\ Sun \& A.\ Vikhlinin}
\affil{Harvard-Smithsonian Center for Astrophysics, 60 Garden St.,
Cambridge, MA 02138;\\ msun@cfa.harvard.edu}

\shorttitle{X-rays from UGC 6697}
\shortauthors{Sun \& Vikhlinin}

\begin{abstract}

We present the result from a \chandra\ observation of an X-ray luminous
starburst galaxy UGC~6697, which is embedded in the northwest hot region
of A1367 (5 - 6 keV). A very sharp X-ray edge ($\sim$ 13 times of surface
brightness jump) at the southeast and a long tail (at least 60 kpc from the
nucleus) at the northwest of the galaxy are detected, as expected if the
galaxy is moving to the southeast. The X-ray edge, at the midway of the
nucleus and the southeast optical disk edge, is also
at the same position where the H$\alpha$ emission is truncated and a radio
sharp edge is observed. The X-ray diffuse
emission is also enhanced at the southeast, implying ram pressure compression.
No extraplanar X-ray component is detected, probably due to the combining
effects of weaker outflow activity than that in nuclear starbursts, and external
confinement plus stripping. The diffuse thermal gas in UGC~6697 has a temperature
of $\sim$ 0.7 keV and a low iron abundance ($\sim$ 0.1 - 0.2 solar).
An X-ray point source (L$_{\rm 0.5-10 keV} \sim 2.8 \times 10^{40}$ ergs s$^{-1}$)
is detected on the nucleus, but not highly absorbed. Three off-center ultraluminous
X-ray sources, all with L$_{\rm 0.5-10 keV} > 10^{40}$ ergs s$^{-1}$, are also
detected. Based on the multi-wavelength data available, we favor that the
interaction between the interstellar medium (ISM) and the ICM plays a major
role to trigger the starburst in UGC~6697.

\end{abstract}


\section{Introduction}

The dynamical and thermal properties of the hot intracluster medium (ICM) have
long been proposed to play vital roles in shaping the properties of
cluster galaxies. One significant question is the effect of the ICM
on the cluster star formation rate (SFR).
After the discovery of the ``Butcher-Oemler effect'' in the medium redshift
clusters (z $\approx$ 0.4; Butcher \& Oemler 1978, 1984), it was
suggested that compression by the ICM for the first-infall galaxies is
the mechanism for star formation in these blue galaxies (e.g., Gunn 1989).
Recent 3D simulations on ram pressure stripping of
spirals infalling into a cluster (Abadi, Moore \& Bower 1999; Quilis,
Moore \& Bower 2000; Vollmer et al. 2001; Schulz \& Struck 2001) have
shown the complexity of this process. Star formation is largely
suppressed in high density ICM regions due to the depletion of the galactic
medium, but SFR can be boosted in less dense ICM regions when the
galactic medium is only compressed but not yet stripped. Recently, Bekki \&
Couch (2003) suggested that strong compression of galactic molecular
clouds due to the high pressure of ICM, either static pressure or ram
pressure, can trigger starburst activity lasting on the order of 10$^{7}$ yr.
The rich cluster Abell~1367 is a good nearby system to study starburst
activity in clusters, where three subgroups containing high fraction of
star-formation galaxies are falling into the cluster core (summarized
by Cortese et al. 2004). In this work, we present the analysis of the X-ray
emission from the most peculiar and the largest galaxy in these infalling
groups, UGC~6697, based on our \chandra\ observations.

UGC 6697 is a starburst galaxy with a SFR of 5 M$_{\odot}$ /
yr (Kennicutt, Bothun, \& Schommer 1984; Donas et al. 1990) and a high IRAS
60 $\mu$m to 100 $\mu$m flux ratio (f$_{60}$/f$_{100}$) of 0.5. The star
formation activity is wide spread on the disk and not concentrated on the
nucleus (Gavazzi et al. 1995; G95 hereafter). The galaxy lies 18$'$
(or 450 kpc) northwest of the X-ray peak of
the rich galaxy cluster A1367 and appears falling into the cluster. In the
optical bands (U, B, V and R; from GOLD Mine, Gavazzi et al. 2003),
it is classified as a peculiar galaxy with
an asymmetrical appearance. However, its morphology in the H band is rather
symmetrical around the nucleus (G95), which implies that the optical asymmetry
is mainly caused by the recent star formation in the disk and dust extinction.
A radio "trail" on the northwest side was detected by Gavazzi \& Jaffe (1987).
HI was found to distribute in the similar asymmetrical way as the optical
light and radio continuum and 70\% of the neutral hydrogen was found in the northwest
side of the galaxy (Sullivan et al. 1981; Dickey \& Gavazzi 1991). G95
summarized more unusual properties of UGC~6697 based on radio, optical
and NIR data: supergiant HII regions, on the periphery of the galaxy, about
10 times more numerous than normal for its luminosity; shocked gas inferred
from strong [O I] and [S II] emission; and 10 times more luminous synchrotron
radiation than in other similar galaxies. Those peculiarities have been
interpreted as due to the dynamical interaction between the ICM and the
galaxy gas, assuming UGC~6697 is moving southeast. However, Gavazzi et al.
(2001) revealed a complicated velocity field of UGC~6697 and suggested
that UGC~6697 is composed of two galaxies interacting.

X-ray emission of the starburst galaxies is a good probe of the mechanical feedback on
galactic scales and the ICM-ISM interaction. Fruscione \& Gavazzi (1990) first
reported X-ray emission from UGC~6697 based on the \einstein\ data. The galaxy
was barely resolved in an old \chandra\ observation as it was $\sim$ 17$'$
from the aim point (Sun \& Murray 2002). A new \chandra\ observation
centered at $\sim 1.5'$ east of UGC~6697 has better resolved its X-ray emission
as presented in this work.

The average velocity of UGC~6697 is 6725 $\pm$ 2 km/s (from NASA/IPAC
Extragalactic Database), 245 km/s larger, in the line of sight, than
the average velocity of the northwest subcluster of A1367 (from Cortese
et al. 2004). In this paper, we adopt the average velocity
of the northwest subcluster to calculate the distance of UGC~6697. Assuming
H$_{0}$ = 70 km s$^{-1}$ Mpc$^{-1}$, $\Omega$$_{\rm M}$=0.3, and
$\Omega_{\rm \Lambda}$=0.7, the luminosity distance of UGC~6697 is
94.4 Mpc, and 1$''$=0.438 kpc. The B band luminosity of UGC~6697
(L$_{\rm B}$) is 5$\times10^{10}$ L$_{\odot}$. Uncertainties quoted
are 1 $\sigma$.

\section{\chandra\ observations}

A 48 ksec \chandra\ observation was performed with the Advanced CCD Imaging
Spectrometer (ACIS) on January 24, 2003. The data analysis is the same as
the one presented in Sun et al. (2004), as the same \chandra\ pointing is
analyzed. The calibration files used correspond to CALDB 2.28 from the
\chandra\ X-ray Center (CXC).

\subsection{The X-ray diffuse emission in UGC~6697}

The \chandra\ count image of UGC~6697 is shown in Fig. 1. The X-ray emission
is enhanced in the southeast part of the galaxy with a sharp edge at the
southeast end. A diffuse tail is observed to the northwest of the galaxy.
We measured the \chandra\ surface brightness (exposure-corrected) along
two slices to trace the tail and the edge (Fig. 1). A brightness jump of
$\sim$ 13 times is observed across the edge. The width of the edge is less
than 1.5$''$ (or 0.7 kpc). The tail extends all the way to the edge of the
S3 chip (or at least 60 kpc from the nucleus).
These features imply the action of ram pressure as the galaxy is
moving to the southeast. The X-ray emission of UGC~6697 appears constrained
to the plane with no extraplanar component generally found in starburst
galaxies in the field or in poor environments (e.g., Strickland et al.
2004). Since the starburst-driven superwind can have a velocity of $\lsim$
1000 km/s, not very smaller than the expected velocity of UGC~6697, the lack of
significant extraplanar component implies that the outflows in UGC~6697
are weaker than that of nuclear starburst (e.g., M82) and can be
easily stripped away by ram pressure.
For its IR luminosity (L$_{\rm IR}$ = 4.3 $\times 10^{10}$
L$_{\odot}$), UGC~6697 is among the X-ray brightest starbursts known
(Strickland et al. 2004), which is likely due to the compression by
ram pressure and static ambient pressure. The X-ray contours,
superposed on the B band optical image, are shown in Fig. 2. The
X-ray edge is well inside the stellar emission of the galaxy, which
is much more symmetric than the X-ray emission since stars are not affected
by the ram pressure.

We also compare the diffuse X-ray emission with the H$\alpha$ emission
(Fig. 2). It had long been noticed that the H$\alpha$
emission truncates interior to the optical continuum emission (G95).
Only when the X-ray emission is well resolved, do we notice
that the truncation of H$\alpha$ emission is at the same position as the
X-ray edge. Within the galaxy, a positive correlation of H$\alpha$
and X-ray emission is observed, which implies that the X-ray emitting
gas is mainly located in the recent star formation regions.

The integrated spectrum of the diffuse emission (Fig. 3) was studied. Since the
system is close to edge-on, the average absorption is likely higher than the
Galactic value (2.2$\times$10$^{20}$ cm$^{-2}$). The average optical
extinction was estimated to be $\sim$ 1 mag by G95, which corresponds
to an absorption excess of $\sim$ 1.8$\times$10$^{21}$ cm$^{-2}$ based
on the averaged relation (with large dispersion) derived by Predehl \&
Schmitt (1995). We present the spectral fits both with absorption
allowed to vary and with it fixed at the Galactic value (Table 1).
Both MEKAL and VMEKAL models were applied. The solar photospheric
abundance table by Anders \& Grevesse (1989) is adopted. VMEKAL model
gives slightly better fits. The gas temperature is $\sim$ 0.7 keV.
The derived absorption is comparable to or not much higher than 
the Galactic value. This is consistent with the bright near-Ultraviolet
(UV) emission from the galaxy (Marcum et al. 2001). The 0.5 - 10 keV
luminosity is 1.0 - 1.2 $\times 10^{41}$ ergs s$^{-1}$. The iron abundance
is low, less than 0.35 solar (90\% higher limit) in the VMEKAL fit.
Adding one more MEKAL component does not improve the fit significantly.
Since the X-ray emission of X-ray binaries is dominated
by the several most luminous ones, the unresolved point source
contribution is small, $<$ 5\% based on the spectral analysis. Low
metallicity measured from the global spectrum of the starburst galaxy
has been reported and discussed before (e.g., Weaver, Heckman \&
Dahlem 2000). Recent work by Fabbiano et al. (2004) on the Antennae
demonstrates that when the data allows spectroscopy in individual regions
to better disentangle the multi-component X-ray gas, solar or over-solar
abundances are derived in active star-formation regions, although
the global spectrum of Antennae only shows a small
abundance. Therefore, the X-ray emitting gas of UGC~6697 is also
very likely multi-phase, which makes the estimate of emission
measure (or gas density) very uncertain, due to the degeneracy between
the abundance and emission measure.
Besides the global temperature, we also examined the hardness ratio
distribution in UGC~6697 but found no significant difference
across the galaxy with the current statistics ($\sim$ 810 counts total
in the 0.7 - 5 keV band, from the diffuse emission of UGC~6697).
Assuming a dimension of 2.5$'\times2.5'\times0.4'$ oblate spheroid and
a best-fit VMEKAL model (Table 1), the
average electron density is 4.6 $\times 10^{-3}$ f$^{-1/2}$ cm$^{-3}$
and the total mass of the detected X-ray emitting gas in UGC~6697 is
3.1 $\times 10^{9}$ f$^{1/2}$ M$_{\odot}$, where f is the filling factor
of the X-ray emitting gas.

The surrounding ICM (projected) has a temperature of $\sim$ 7 - 8 times the
X-ray gas in UGC~6697 (Table 1). We argue that UGC~6697 is not in a low
density region along the line of sight because the observed sharp X-ray
edge indicates a large ram pressure and the northwest subcluster is not
very massive (velocity dispersion of $\sim$ 770 km/s, Cortese et al. 2004).
The $\beta$-model fit to the northwest subcluster by
Donnelly et al. (1998) gives an electron density of $\approx$
5.4 $\times 10^{-4}$ cm$^{-3}$ at the projected position of UGC~6697,
and $> 3.6 \times 10^{-4}$ cm$^{-3}$ if the galaxy is within 10$'$
(or 263 kpc) of its projected position along the line of sight. 
In this work, we take a typical value of 4.5 $\times 10^{-4}$ cm$^{-3}$.
The ICM thermal pressure is then 7.9 $\times 10^{-12}$ dyn cm$^{-2}$
for an ICM temperature of 5.7 keV (Table 1). For comparison, the
ram pressure is 8.7 $\times 10^{-12}$ (v$_{\rm gal}$ / 1000 km/s)$^{2}$
dyn cm$^{-2}$.

\subsection{X-ray point sources in UGC 6697}

Four X-ray point sources (P1 - P4) were detected in UGC~6697 (Fig. 2).
Spectral fits were made to the integrated spectrum of four sources
(Table 1), which shows that they indeed are hard X-ray sources. Their
properties are listed in Table 2.
P2 and P3 are a little extended but that extension may come from the
nearby diffuse emission as all four sources are hard X-ray sources.
We plot the positions of the four sources on the \hst\ UV image (Fig. 4).
The position of P3 is consistent with the proposed nuclear position
of UGC~6697 (the north blue knot A4 in Gavazzi et al. 2001), which is
also the peak in the NIR H band and radio emission. No significant
absorption excess ($> 10^{21}$ cm$^{-2}$) is found to P3. Its
0.5 - 10 keV luminosity is 2.6 - 3.0 $\times10^{40}$ ergs s$^{-1}$.
P1, P2 and P4 have 0.5 - 10 keV luminosity of 1.0 - 2.2 $\times 10^{40}$
ergs s$^{-1}$. P2 is in a H$\alpha$ and UV bright region. P1 is
positionally coincident with a knot in the \hst\ image, while P4 is
$\sim$ 0.3 kpc west to a bright knot. \hst\ I-band image is then examined.
Both knots are very blue, actually comparable or bluer than the
bright complex 7$''$ southeast of P4 (or region c2 in G95). Thus,
P1 and P4 may be associated with young star clusters, similar to other
Ultraluminous X-ray sources detected in starburst galaxies (e.g.,
Fabbiano, Zezas \& Murray 2001; Kaaret et al. 2004). The summed
spectrum of P1, P2 and P4 is consistent with that of high-mass X-ray
binary (HMXB) (Table 1).
The total 2 - 10 keV luminosity of these three ultraluminous X-ray
sources (ULXs) is $\sim 2.9\times10^{40}$ ergs s$^{-1}$, which
is close to the predicted total X-ray luminosity of HMXBs in UGC~6697
(3.3$\times10^{40}$ ergs s$^{-1}$) based on the L$_{\rm X}$ - SFR
relation (Grimm, Gilfanov \& Sunyaev 2003). This again demonstrates
that these ULXs are tightly related to the starburst.
Source confusion in UGC~6697 is not expected to affect the number
of $>$ 10$^{40}$ ergs s$^{-1}$ sources much because: 1) the X-ray
emission is dominated by the most luminous sources;
2) the star formation in UGC~6697 is wide spread and not concentrated
in the nuclear region; 3) on average, less than one bright X-ray source
is expected in or close to a star cluster (e.g., Kaaret et al. 2004).
We also performed the K-S test to the X-ray light curves of these
point sources but found no significant variations.

\subsection{X-ray emission from 2MASX J11434983+1958343}

Diffuse X-ray emission is also detected from the nearby dwarf spiral galaxy
2MASX J11434983+1958343 to the north of UGC~6697 (Fig. 2 and 4). This dwarf
galaxy is H$\alpha$ bright (Fig. 2) and is only $\sim$ 12 kpc to the
plane of UGC~6697 at the projected position. Assuming a massive halo of
2$\times10^{12}$ M$_{\odot}$ within 20 kpc of UGC~6697, the escaping
velocity at 20 kpc is 930 km/s, somewhat larger than the velocity
difference between the dwarf galaxy and UGC~6697 ($\sim$ 800 km/s).
Therefore, the dwarf galaxy is still possible to be bound with UGC~6697.
Its X-ray emission is soft and the estimated 0.5 - 10 keV luminosity
is 4 $\times 10^{39}$ ergs s$^{-1}$. The X-ray emission may come from
the gas heated by active star formation (probably triggered by the tidal
force of UGC~6697) in the dwarf galaxy.

\section{Discussion}

\subsection{The stripping of UGC~6697's X-ray emitting gas}

This \chandra\ observation presents strong evidence for the interaction
between the X-ray gas of UGC~6697 and the surrounding ICM, caused
primarily by the motion of UGC~6697 towards the southeast, relative to the
surrounding ICM. Although a clear surface brightness edge is found in
the X-ray data, the velocity of the galaxy cannot be well constrained
from the X-ray data because of the unknown geometry of the galactic X-ray
emitting gas and the large uncertainty in the emission measure
from the poor determination of abundance. If we simply assume the galaxy
is viewed edge-on and the X-ray gas is in a disk with a filling factor of
unity, the average electron density at the southeast end is 1 - 2
$\times 10^{-2}$ cm$^{-3}$, using the emission measured derived from the
VMEKAL fit. This yields an
internal thermal pressure 2.7 - 5.4 times the ICM thermal pressure.
The X-ray ISM density just inside the edge is expected to be smaller
than the average value. Thus, the velocity of UGC~6697 through the ICM is $<$
1.2 - 2.0 $\times 10^{3}$ km/s for pressure balance.
UGC~6697's direction of motion should
be close to its disk plane, which makes the ram pressure stripping
very difficult. To the first order, we can apply the simple criterion for ram
pressure stripping given in Gunn \& Gott (1972). Even if we assume a small 
angle of 20 degrees between the disk plane and direction of motion, the
critical velocity of UGC~6697 is $\sim$ 3000 km/s (larger for smaller
angle) for its X-ray ISM to be stripped out of the plane. Thus, although
ram pressure can explain the southeast edge, it is not strong enough
in the current environment to fully strip the X-ray gas and other
galactic medium. As pointed out by Nulsen (1982), stripping by transport
processes may be much more important than ram pressure stripping.
The typical mass loss rate due to Kelvin-Helmholtz (K-H) instability is
(Nulsen 1982):

\begin{eqnarray}
\dot{M}_{\rm KH} &\approx& \pi r^2 \rho_{\rm ICM} {\rm v_{gal}} \nonumber \\
&=& 16.5 (\frac{n_{\rm e, ICM}}{4.5 \times 10^{-4} {\rm cm}^{-3}}) (\frac{r}{20 {\rm kpc}})^2
(\frac{{\rm v}_{\rm gal}}{1000 {\rm km/s}}) {\rm M_{\odot} / yr}
\end{eqnarray}

The X-ray gas mass in the tail is $\sim 10^{8}$ M$_{\odot}$, which can be
explained by the K-H instability acting for a time duration shorter than or
comparable  to the lifetime of the starburst ($\sim 10^{7}$ yr).
The mass loss due to thermal conduction is $\sim$ 3.5 times larger
(Nulsen 1982), but the efficiency of thermal conduction is uncertain
due to the presence of magnetic fields implied by the strong radio
emission. The field lines may have been stretched and entangled at
the boundary in the stripping process to largely suppress the
thermal conduction, as what was found for the cool coronae in hot
clusters (e.g., Sun et al. 2004). Besides being the gas removed by
the K-H instability at the front and the sides, the X-ray tail of
UGC~6697 may have minor contribution from the gas heated by active star
formation in the tail (implied by its blue color and the event of
the Type II SN 1986C there). With a similar hardness ratio as the
X-ray gas in bright regions, the X-ray tail is not likely to
be the concentrated ICM emission by e.g., Bondi accretion, as the
ICM temperature is high (5 - 6 keV) and the galaxy velocity is likely
large. The current SFR is $\sim$ 5 M$_{\odot}$
yr$^{-1}$, which will likely last for about several 10$^{7}$ yr.
Since massive stars only have lifetimes on the order of 10$^{6}$ yr,
the mass loss rate from stars due to stellar winds and SN explosions
is an order of magnitude higher than the SFR and can replenish for the
large mass lost due to transport processes and ram pressure stripping.

The southeast edge is very sharp.
For the measured gas temperature and density inside and outside of the
ICM-ISM boundary, the mean free path of electrons in the hot ICM
is $\sim$ 22 kpc. The mean free path of electrons from the hot ICM to the
cool ISM is $\sim$ 4.3 (n$_{\rm e, ISM}$/10$^{-3}$ cm$^{-3}$)$^{-1}$ kpc, while
the mean free path of electrons from the cool ISM to the hot ICM is
$\sim$ 4.7 kpc (Spitzer 1962). These values are much larger than the upper
limit on the width of the edge ($\sim$ 0.7 kpc). Thus, the particle diffusion
across the boundary has to be suppressed by at least 6 times,
similar to the situation in cluster cold fronts (Vikhlinin et al. 2001).

The X-ray tail of UGC~6697 is only the second known for a spiral in
a rich cluster, next to the galaxy C153 in the 3.2 keV cluster A2125 (Wang
et al. 2004). Because of its proximity (compared to A2125 at z=0.247),
UGC~6697 is the best example known for the stripping of a spiral's
X-ray gas in a rich cluster (e.g., no gas temperature measured for
the X-ray gas of C153). This phenomenon should be rare in the
elliptical-dominated rich clusters, as high star formation activity, which only lasts
$\sim 10^{7}$ yr, is required to replenish the gas loss by fast stripping.
In the case of UGC~6697, the edge-on stripping also makes it easier
to retain the galactic X-ray gas.

\subsection{A starburst triggered by the ICM-ISM interaction?}

Although ram pressure is not large enough to deplete the galactic
gas (even only the X-ray gas), it can still significantly impact the
evolution of UGC~6697. It has been suggested that the starburst in
UGC~6697 was triggered by the compression of the hot ICM (e.g. G95).
This is an interesting question to address, since the mechanism to
cause the starburst activity in UGC~6697 may be related to the
Butcher-Oemler effect at the medium redshift. The \chandra\ data clearly
show a leading edge produced by the ram pressure at the front.
The positional coincidence of the X-ray
edge and the H$\alpha$ truncation position implies their connection.
The H$\alpha$ emission is mainly from $\sim$ 10$^{4}$ K gas heated
by the nearby young massive stars and is sensitive to stellar population
with ages of $< 10^{7}$ yr. A definite edge, well inside the optical
continuum image, implies the truncation of star formation activity upstream
from the edge. This may be best explained by the action of ram pressure,
if the starburst activity in this galaxy is triggered by the interaction
with the ICM. As Strickland et al. (2000) show for the nearby starburst
galaxy NGC~253, the X-ray emission is mainly from the regions where
the fast SB-driven wind interacts with the denser ambient
ISM. Since the mass loss rate from stars due
to wind and SN explosions is high and the starburst duration is short,
the X-ray diffuse emission generally traces the star formation regions,
as shown by the general correlation between X-ray emission and
H$\alpha$ emission in starburst galaxies. Thus, both the X-ray edge  
and the H$\alpha$ edge may reflect the current front of active star
formation. Upstream from the front, the galaxy is still bright in
U band (G95). Since the UV flux is sensitive to the stellar population
of age $\approx 10^{8}$ yr, the stellar population upstream from the
front is indeed older. Most of ISM there has been stripped so the current
star formation activity is weaker. There is also a significant
radio edge at the position of the H$\alpha$ edge (G95). The synchrotron
radiation there may be boosted by the compression of magnetic field
due to the ram pressure. Thus, the three edges (radio, H$\alpha$ and
X-ray) may all correspond to the same front where the star formation is
currently truncated outside due to stripping. The starburst
activity at the front and inside the front is triggered by the interaction
with the ICM.
The properties of UGC~6697's environment match those assumed in the
simulation by Bekki \& Couch (2003), e.g., ICM density not too high
to cause too much stripping. The SFR predicted in their
simulation is 0.1 - 0.6 M$_{\odot}$ yr$^{-1}$ for $\sim$ 10$^{7}$ yr
for a single cloud. This can easily explain the total SFR
of UGC~6697. UGC~6697 may not be a
perfect edge-on galaxy from its H$\alpha$ emission and the direction
of motion of the galaxy may not align exactly with the disk plane of
the galaxy. Thus, a velocity component to the north could explain the
largely enhanced H$\alpha$ and radio emission there compared to the
south.

However, Gavazzi et al. (2001) proposed that UGC~6697 is composed of
two interacting galaxies. The smaller one lies behind the main galaxy
and the northwest tail is a tidal tail of the main galaxy. In this
scenario, UGC~6697 is like an Antennae galaxy in a specific configuration
viewed edge-on, i.e., two galaxy planes and our line of sight are
nearly in the same plane. While the idea is attractive to explain
the complex velocity field, there is no features detected in either
the \chandra\ image or the hardness ratio map that indicates the
interaction of two galaxies. There is also no bright X-ray enhancement
around the nucleus (P3), as expected from the nuclear starburst
induced by tidal interaction. The H-band image (G95) also looks
very normal for single galaxy. Nevertheless, the peculiarities
of UGC~6697 cannot be all explained by ram pressure stripping.
The warp at the southeast end of the stellar disk is a clear
indication of tidal interaction by the companion galaxies (visible
in Fig. 2 and Fig. 4), if not by the proposed second galaxy (Gavazzi et al. 2001).
To sum up, we consider that the ICM-ISM interaction plays a
major effect on the wide-spread star formation in UGC~6697, but tidal
effects need to be carefully examined, from the multi-wavelength
data and simulations.

\section{Conclusion}

In this work, we present clear evidence, a sharp leading edge and
a long X-ray tail (at least 60 kpc from the nucleus), for the
interaction between the $\sim$ 0.7 keV X-ray gas of the starburst
galaxy UGC~6697 and its surrounding hot ICM of A1367 ($\sim$ 5.7 keV)
from the \chandra\ observation. This rare case is the best-known
example of the X-ray gas of a spiral being stripped by the ICM
of a rich cluster and only the second known.
The X-ray edge is caused by ram pressure
as the galaxy falls through the hot ICM. The X-ray emission of
UGC~6697, most likely from the gas heated by SN explosions, appears
confined to the plane and no extraplanar component is detected.
The derived iron abundance from the global spectrum is low ($<$ 0.35
solar for the 90\% higher limit).
Apart from an X-ray point source detected from the nucleus, three
ULXs (L$_{\rm 0.5 - 10 keV} > 10^{40}$ ergs s$^{-1}$) are detected
with hard spectra. The estimated total X-ray luminosity of
HMXBs and the SFR of UGC~6697 follow the L$_{\rm X}$ - SFR relation
by Grimm et al. (2003), which implies that these luminous sources are
the products of the ongoing starburst.
The X-ray edge is at the same position as the H$\alpha$ edge and the radio
edge-like feature. We suggest that these three edges all correspond to
the same front. Star formation is truncated outside of the front and the
starburst inside of the front is likely triggered by the ICM-ISM interaction.

\acknowledgments

We thank W. Forman, C. Jones and P. Nulsen for valuable comments. 
We also thank the referee G. Gavazzi for prompt comments. We are
grateful to G. Gavazzi for providing us the optical data of UGC~6697. This
research has made use of the GOLD Mine Database, operated by the Universita'
degli Studi di Milano- Bicocca. The financial support for this work was
provided by \chandra\ grant GO3-4161X.

\begin{table}
\begin{center}
\caption{The Spectral Fits}
{\scriptsize
\begin{tabular}{ccccccc} \hline \hline
 & Model & Absorption$^{\rm a}$ & Parameters$^{\rm b}$ & $\chi^{2}$/d.o.f. & L$_{\rm 0.5 - 10 keV}^{\rm c}$ & L$_{\rm bol}^{\rm c}$ \\ \hline
UGC~6697 diffuse emission$^{\rm d}$ & MEKAL & (0.22) & T=0.77$^{+0.05}_{-0.06}$, Z=0.07$^{+0.03}_{-0.02}$ & 29.4/21 & 0.93 & 1.95 \\
     &  & 0.67$^{+0.70}_{-0.33}$ & T=0.66$^{+0.11}_{-0.06}$, Z=0.05$^{+0.02}_{-0.01}$ & 28.6/20 & 1.24 & 2.52 \\
     & VMEKAL & (0.22) & T=0.80$^{+0.06}_{-0.10}$, O=1.1$^{+0.7}_{-0.5}$, Si=0.2$^{+0.4}_{-0.2}$, Fe=0.17$^{+0.09}_{-0.06}$ & 23.3/17 & 0.96 & 1.63 \\ 
     &  & 0.69$^{+0.73}_{-0.49}$ & T=0.74$^{+0.09}_{-0.16}$, O=1.0$^{+0.9}_{-0.61}$, Si=0.2$^{+0.3}_{-0.2}$, Fe=0.12$^{+0.08}_{-0.03}$ & 22.3/16 & 1.26 & 2.28 \\
Surrounding ICM$^{\rm e}$ & MEKAL & (0.22) & T=5.66$^{+0.70}_{-0.55}$, Z=0.04$^{+0.18}_{-0.04}$ & 100.8/90 & - & - \\
Four point sources$^{\rm f}$ & Power law & (0.22) & $\Gamma$=1.60$^{+0.10}_{-0.09}$ & 16.0/24 & 0.77 & 2.73 \\
     &  & 0.82$^{+0.66}_{-0.50}$ & $\Gamma$=1.77$^{+0.22}_{-0.16}$ & 14.5/23 & 0.77 & 2.34 \\
     & Bremsstrahlung & (0.22) & T=6.7$^{+2.7}_{-1.7}$ & 14.9/24 & 0.68 & 0.95 \\
     &  & 0.24$^{+0.66}_{-0.50}$ & T=6.6$^{+3.9}_{-2.1}$ & 14.9/23 & 0.68 & 0.95 \\
Three point sources$^{\rm g}$ & Power law & (0.22) & $\Gamma$=1.68$\pm$0.15 & 26.7/24 & 0.47 & 1.41 \\
     &  & 1.65$^{+0.93}_{-0.76}$ & $\Gamma$=1.98$^{+0.27}_{-0.19}$ & 23.3/23 & 0.50 & 1.51 \\
     & Bremsstrahlung & (0.22) & T=6.5$^{+3.7}_{-1.9}$ & 24.9/24 & 0.42 & 0.56 \\
     &  & 0.76$^{+0.69}_{-0.54}$ & T=4.7$^{+2.8}_{-1.5}$ & 24.0/23 & 0.41 & 0.56 \\
\hline \hline
\end{tabular}
\begin{flushleft}
\leftskip 35pt
$^{\rm a}$ The absorption column, fixed (in brackets, the Galactic value) or free, is in unit of 10$^{21}$ cm$^{-2}$. \\
$^{\rm b}$ The temperature unit is keV and abundances are in solar abundances. \\
$^{\rm c}$ The unit of X-ray luminosity is 10$^{41}$ ergs s$^{-1}$. \\
$^{\rm d}$ Diffuse emission from the elliptical region in Fig. 1\\
$^{\rm e}$ Cluster diffuse emission within 3$'\times2'$ (semimajor axis $\times$
semiminor axis) ellipse centered on UGC~6697 (sources excluded) \\
$^{\rm f}$ Sum spectrum of four X-ray point sources in UGC~6697\\
$^{\rm g}$ Sum spectrum of three X-ray point sources (excluding the nucleus source P3)\\
\end{flushleft}}
\end{center}
\end{table}

\begin{figure}
\vspace{-10cm}
   \centerline{\includegraphics[height=1.4\linewidth]{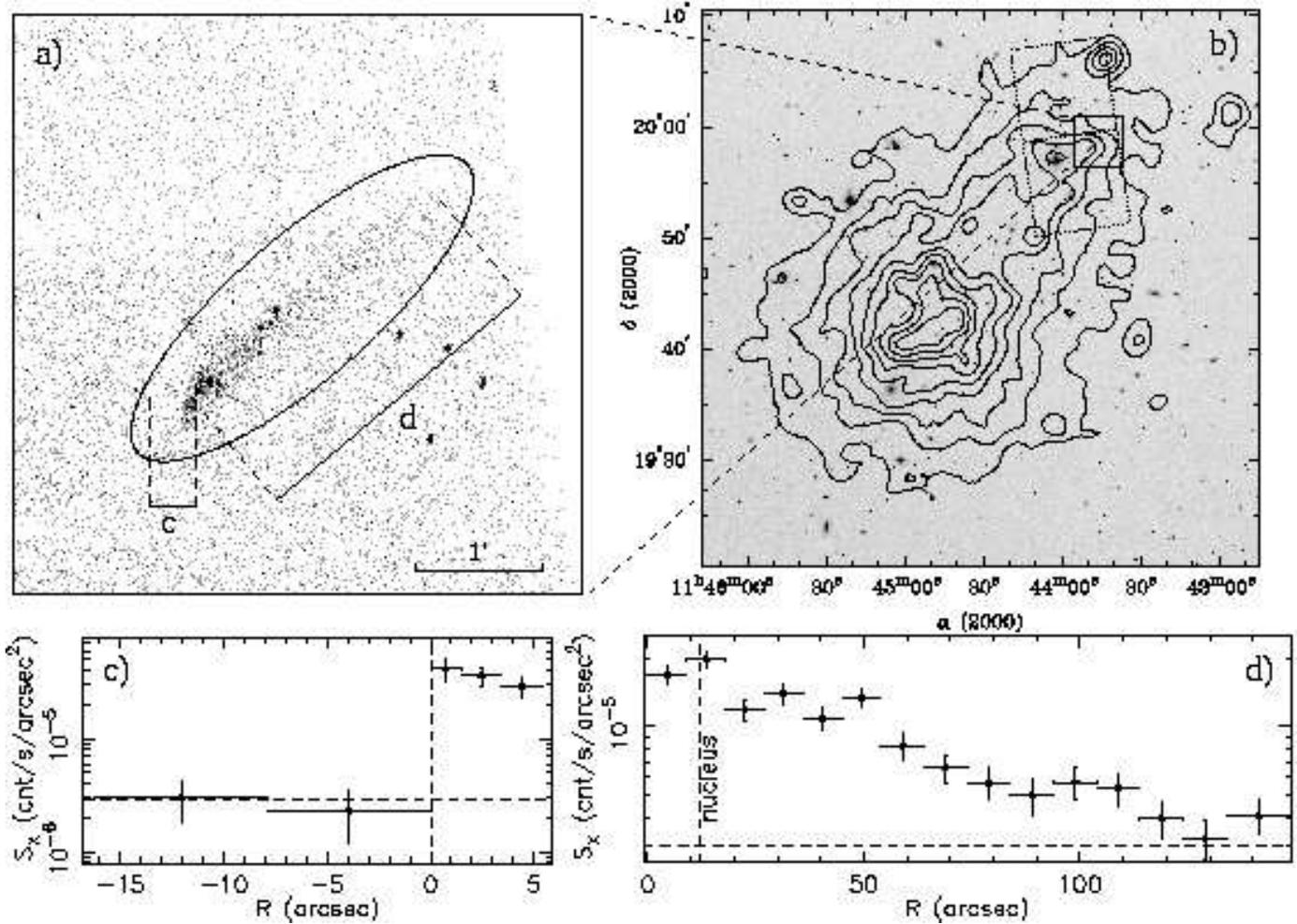}}
  \caption{a): 0.5 - 2 keV \chandra\ count image of UGC~6697. A sharp
surface brightness edge at the southeast end and a long tail to the
northwest are significant. The detected X-ray emission is confined to
the galactic plane. The selected contrast of the image does not show the
point sources well, but four X-ray point sources are detected in UGC~6697
(shown in Fig. 2). The surface brightness is measured along the two
narrow slices (shown in c and d). The ellipse is the region where the
global spectrum of UGC~6697 diffuse emission was extracted (see $\S$2.2).
b): \rosat\ contours of A1367 superposed on the DSS image. The two
squares (dotted lines) show the S2 (north) and S3 (south) CCD fields.
The small rectangle (solid lines) is the enlarged region (on the left).
UGC~6697 is in a subcluster that is merging with the southeast subcluster.
c): The surface brightness profile across the edge at the south end.
Vertical box regions (8.36$''$ width) are used to measure the surface
brightness. A brightness jump of $\sim$ 13 times is observed across the
edge. The horizontal dashed line represents the background level (also
in d). The width of the edge is less than 1.5$''$ or 0.7 kpc. d): The
surface brightness profile (excluding point sources) along the galactic
plane of UGC~6697 (20$''$ width adopted). The tail extends to the edge
of the S3 chip.
   \label{fig:img:smo}}
\end{figure}

\begin{table}
\begin{center}
\caption{The X-ray point sources in UGC~6697}
{\scriptsize
\begin{tabular}{cccccc} \hline \hline
 & RA & decl. & counts & hardness ratio & note \\
 & (2000) & (2000) & (0.7 - 5 keV) & (1.4 - 5 keV / 0.7 - 1.4 keV) & \\ \hline

P1 & 11:43:49.53 & 19:57:49.8 & 43$\pm$7 & 1.02 $\pm$ 0.37 & in a star cluster? \\
P2 & 11:43:49.46 & 19:58:03.1 & 55$\pm$10 & 1.21 $\pm$ 0.36 & \\
P3 & 11:43:49.12 & 19:58:07.3 & 110$\pm$11 & 1.16 $\pm$ 0.28 & UGC~6697 nucleus \\
P4 & 11:43:46.86 & 19:58:40.2 & 98$\pm$10 & 0.80 $\pm$ 0.20 & close to a star cluster? \\

\hline \hline
\end{tabular}}
\end{center}
\end{table}

\begin{figure}
\vspace{-10cm}
  \centerline{\includegraphics[height=1.35\linewidth]{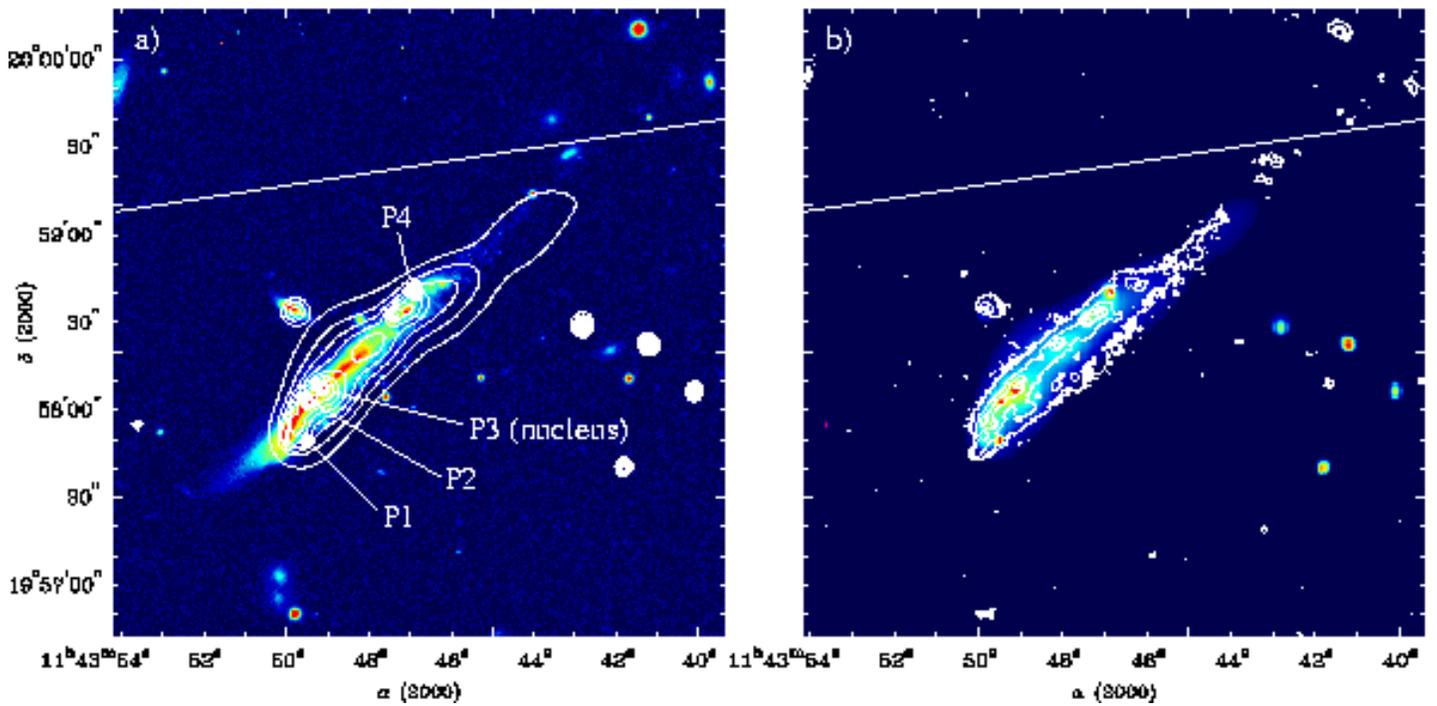}}
  \caption{a): \chandra\ contours (from the exposure-corrected, and adaptively
smoothed image by CSMOOTH)
superposed on the B-band image of UGC 6697. Four X-ray point sources are
marked. The contour levels increase by a factor of $\sqrt{2}$ from the
outermost one (10$^{-3}$ counts/ksec/arcsec$^{-2}$). The solid white line
is the S3 chip boundary (also in b).
The X-ray sharp edge is well inside the optical light. The X-ray emission
is more asymmetric than the optical light (P3 is the proposed nucleus),
as the edge is only 25$''$ (or 11 kpc) from P3 but the tail extends to the
edge of the S3 chip (140$''$ or 61 kpc from P3, see Fig. 1).
b): The H$\alpha$ contours of UGC~6697 superposed on the \chandra\ color
image in the same field of a. The H$\alpha$ emission truncates at the same
position where the X-ray sharp edge is located. A clumpy H$\alpha$ tail to
the northwest, at the same position of the X-ray tail, is also observed.
There is a general correlation between the H$\alpha$ emission and the
diffuse X-ray emission.
   \label{fig:img:smo}}
\end{figure}
\clearpage

\begin{figure}
  \centerline{\includegraphics[height=0.5\linewidth,angle=270]{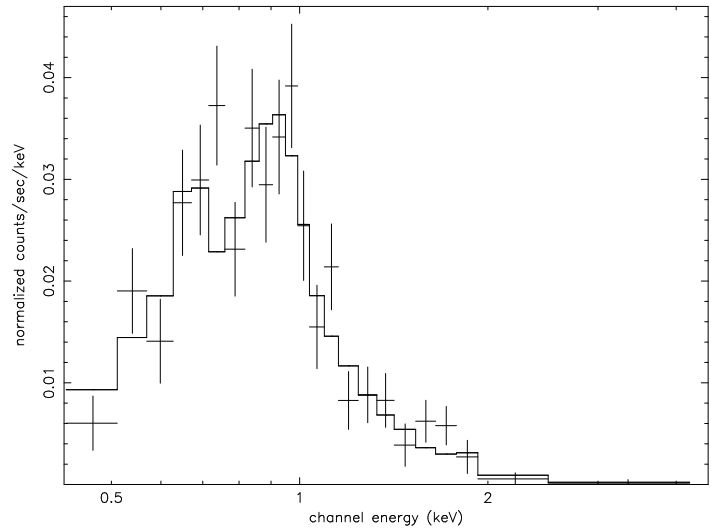}}
  \caption{\chandra\ spectrum of the diffuse emission in UGC~6697 with the
best-fit VMEKAL model. The blend of emission lines can
be seen from 0.7 keV to 1 keV.
   \label{fig:img:smo}}
\end{figure}

\begin{figure}
  \centerline{\includegraphics[height=0.9\linewidth,angle=270]{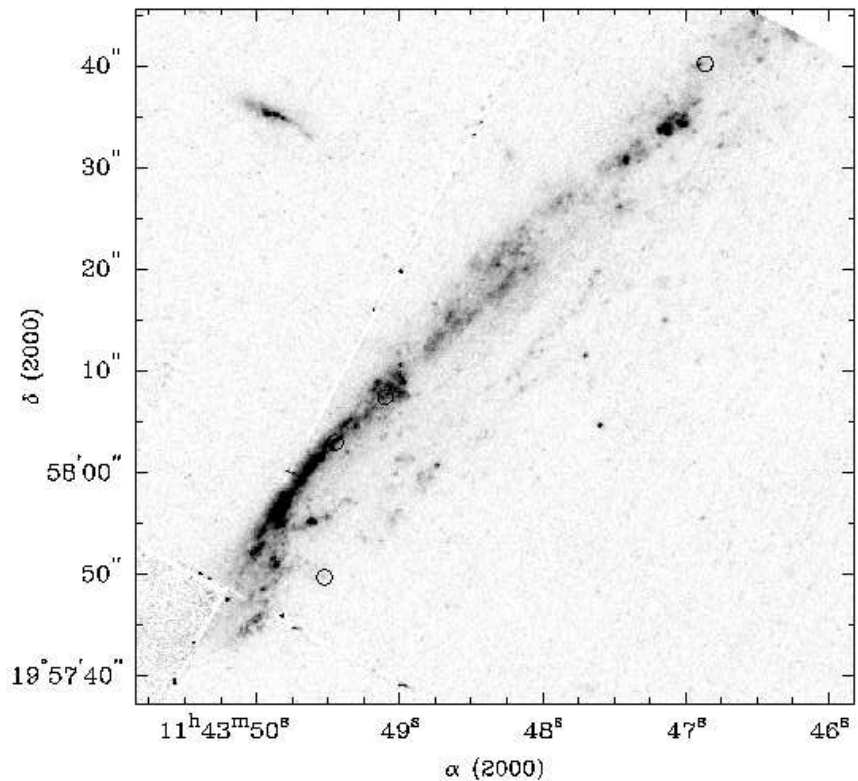}}
  \caption{The positions of the four X-ray point sources superposed on the \hst\ UV
image (F300W filter) (P1 - P4, see Fig. 2). P3 is positionally coincident with the
proposed nucleus of UGC~6697. P1 and P4 may be associated with star clusters.
   \label{fig:img:smo}}
\end{figure}

\end{document}